\documentclass[,trackchanges, twocolumn]{aastex701}

\newcommand{\mytilde}{\raise.19ex\hbox{$\scriptstyle\sim$}}

\begin{document}

%\title{The Discovery of the Highest Redshift Strong Lensing Galaxy Cluster: JWST Reveals Giant Arcs from a Highly Concentrated Cluster at Cosmic Noon}
%\title{JWST Discovery of the Highest-Redshift Strong-Lensing Galaxy Cluster: Giant Arcs from the Highly Concentrated Core of XLSSC 122 at Cosmic Noon}
%\title{The Discovery of Giant Arcs in the Highest-Redshift (z = 1.98) Cluster to Exhibit Strong Lensing: A JWST View of an Exceptionally Concentrated Core}
\title{JWST Discovery of Strong Lensing from a Galaxy Cluster at Cosmic Noon: Giant Arcs and a Highly Concentrated Core of XLSSC 122}

\author[0000-0002-4462-0709]{Kyle Finner}
\affiliation{IPAC, California Institute of Technology, 1200 E California Blvd., Pasadena, CA 91125, USA}
\email[show]{kfinner@caltech.edu}  

\author[0000-0001-7148-6915]{Sangjun Cha}
\affiliation{Department of Astronomy, Yonsei University, 50 Yonsei-ro, Seoul 03722, Korea}
\email[]{sang6199@yonsei.ac.kr} 
\author[0009-0009-4086-7665]{Zachary P. Scofield}
\affiliation{Department of Astronomy, Yonsei University, 50 Yonsei-ro, Seoul 03722, Korea}
\email[]{zpscofield15@gmail.com} 

\author[0000-0002-5751-3697]{M. James Jee}
\affiliation{Department of Astronomy, Yonsei University, 50 Yonsei-ro, Seoul 03722, Korea}
\affiliation{Department of Physics and Astronomy, University of California, Davis, One Shields Avenue, Davis, CA 95616, USA}
\email[]{mkjee@yonsei.ac.kr} 

\author[0000-0001-8792-3091]{Yu-heng Lin}
\affiliation{IPAC, California Institute of Technology, 1200 E California Blvd., Pasadena, CA 91125, USA}
\email[]{ianlin@ipac.caltech.edu}

\author[0000-0001-9139-5455]{Hyungjin Joo}
\affiliation{Department of Astronomy, Yonsei University, 50 Yonsei-ro, Seoul 03722, Korea}
\email[]{gudwls4478@gmail.com}

\author[0009-0007-7093-1758]{Hyosun Park}
\affiliation{Department of Astronomy, Yonsei University, 50 Yonsei-ro, Seoul 03722, Korea}
\email[]{hyosun.park@yonsei.ac.kr}

\author[0000-0002-8512-1404]{Takahiro Morishita}
\affiliation{IPAC, California Institute of Technology, 1200 E California Blvd., Pasadena, CA 91125, USA}
\email[]{takahiro@ipac.caltech.edu}

\author[0000-0002-9382-9832]{Andreas Faisst}
\affiliation{IPAC, California Institute of Technology, 1200 E California Blvd., Pasadena, CA 91125, USA}
\email[]{afaisst@caltech.edu}

\author[0000-0003-1954-5046]{Bomee Lee}
\affiliation{Korea Astronomy and Space Science Institute, 776 Daedeokdae-ro, Yuseong-gu, Daejeon 34055, Korea}
\email[]{bomee@kasi.re.kr}

\author[0000-0002-7964-6749]{Wuji Wang}
\affiliation{IPAC, California Institute of Technology, 1200 E California Blvd., Pasadena, CA 91125, USA}
\email[]{wujiwang@caltech.edu}

\author[0000-0001-7583-0621]{Ranga-Ram Chary}
\affiliation{University of California, Los Angeles, CA 90095-1562, USA}
\email[]{rchary@ucla.edu}

\begin{abstract}

%In the standard cosmological model, Lambda Cold Dark Matter ($\Lambda \mathrm{CDM}$), structure grows hierarchically with the most massive dark matter halos forming late and requiring billions of years to become centrally concentrated. It then follows that galaxy clusters in the early universe are dynamically young and less concentrated. 
Our observations with the James Webb Space Telescope have made the remarkable discovery of strong gravitational lensing arcs from XLSSC~122 ($z=1.98$) - setting the record for the most distant galaxy cluster that exhibits strong lensing. The discovery of giant arcs enables a strong-lensing analysis and a measurement of the concentration of the dark matter halo. We perform a strong-lensing analysis of the cluster and measure the radial projected mass density profile. Our measurements reveal an exceptionally high concentration in the core of XLSSC~122. A Navarro--Frenk--White profile fit to the inner 100 kpc estimates the concentration to be $6.3\pm0.5$. The high concentration of XLSSC~122 contributes to the emerging picture that massive structure formation in the early universe may proceed more rapidly than standard models suggest. We estimate the mass within 100 kpc to be $M$($R<$100 kpc) = $6.5\pm0.7\times10^{13}$ M$_\odot$. Our mosaic images are made public at \href{https://kylefinner.github.io/xlssc122}{JWST mosaics}\footnote{https://kylefinner.github.io/xlssc122}.

\end{abstract}



\section{Introduction} 

The standard cosmological model, Lambda Cold Dark Matter ($\Lambda$CDM), posits
a hierarchical formation of structure, where the most massive dark matter halos
form late and require billions of years to become centrally concentrated. This
paradigm implies that galaxy clusters in the early universe are dynamically young
and less concentrated. Observations with the James Webb Space Telescope (JWST) have challenged the $\Lambda$CDM model by discovering an unexpected abundance of extremely massive galaxies at early epochs of the universe \citep{2023harikane, 2024finkelstein, 2024casey, 2024robertson, 2024donnan}. Spectroscopic measurements confirmed that many of the galaxies have a high redshift $z\gtrsim9$ \citep{2023wang,  2024carniani, 2024fujimoto, 2024castellano} with the most distant galaxy spectroscopically confirmed at $z=14.44\pm0.02$ \citep{2025naidu}. The overabundance of high-z galaxies has invigorated early-galaxy formation theories of galaxy evolution. Significant  contribution from active galactic nuclei has been ruled out \citep{2024carniani}, but dust content, increased star-formation efficiency, or a top-heavy initial mass function have been proposed as possible solutions. Non-standard cosmologies, such as early dark energy \citep[e.g.,][]{2024shen}, have also been proposed to resolve the overabundance of massive galaxies in the early universe. 

\begin{figure*}[t!]
    \centering
    \includegraphics[width=\linewidth]{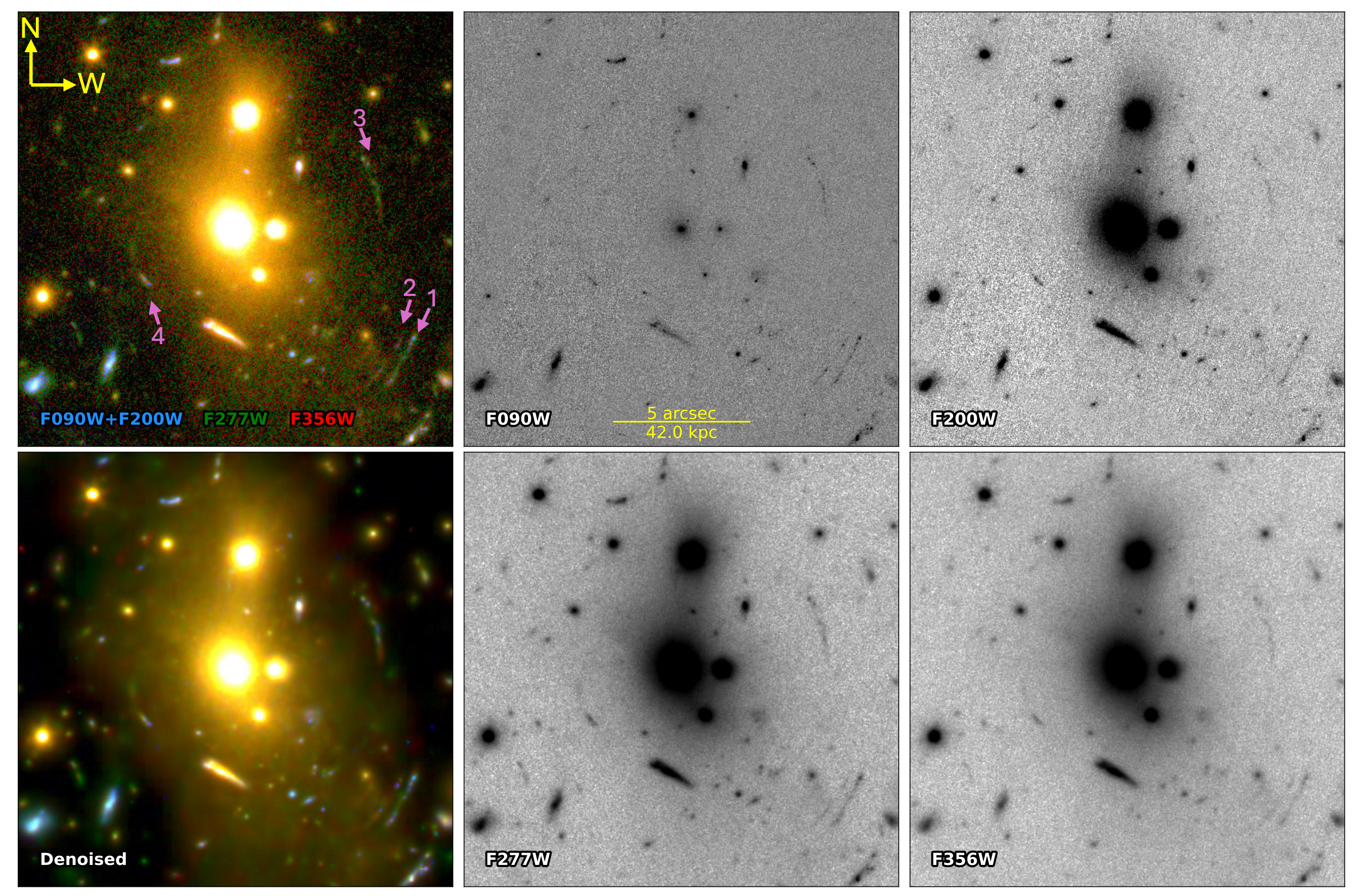}
    \caption{The left panels show the color-composite image (top) and denoised color-composite image (bottom) of the core of galaxy cluster XLSSC~122 with RGB colors from F356W, F200W+F277W, and F090W, respectively. The denoised image was produced using a Transformer-based reconstruction method \citep{2024park, 2025scofield}. The middle and right panels show the log-scale, single-filter images as labeled. The cluster galaxies are much fainter in the F090W filter because, at $z = 1.98$, the filter probes rest-frame wavelengths below the 4000~\AA~break. The strong-lensing arcs are labeled with purple arrows in the top-right image and are visible in all filters. The intracluster light is prominent in the redder filters and stretches from the BCG to the southeast H. Joo et al. (in prep.).}
    \label{fig:xlssc122_image}
\end{figure*}

The galaxy cluster XLSSC~122 ($z=1.98$) has been at the forefront of observational cosmology since its discovery by the \textit{XMM-Newton} X-ray telescope in the 25 sq. deg. XXL Survey \citep{2006pierre, 2016pierre, 2013willis} and the characterization of its gas properties \citep{2014mantz, 2018mantz}. The fame that XLSSC~122 has garnered is well-deserved, with each turn bringing an exciting discovery. Spectroscopic observations with the Hubble Space Telescope (HST) determined that the cluster has a redshift $z=1.98$ \citep{2020willis}, which equates to a look-back time of 10.2 billion years or about 3.3 billion years after the Big Bang. Remarkably, the galaxies in the cluster have characteristics that indicate an old population of stars that started forming when the Universe was only 370 million years old \citep{2020willis}. This discovery places XLSSC~122 as a mature cluster at cosmic noon -- the stage of the Universe where galaxy clusters first appear. Subsequent studies confirmed its status as a mature cluster \citep{2021noordeh, 2022trudeau}. Furthermore, \cite{2021noordeh} found the cluster to host an excess of `bulge-like' quiescent galaxies, which suggests an accelerated size evolution relative to field galaxies.

A key to understanding the formation and evolution of galaxy clusters is characterizing their dark matter, which makes up approximately 85\% of their mass. \cite{2025kim} performed the first weak-lensing analysis of XLSSC~122 with HST observations. They report a $S/N=3.2\sigma$ detection of the mass peak and $M_{200c}=3.3\pm1.8\times 10^{14}$ M$_\odot$. In this work, we present JWST NIRCam observations of XLSSC~122 and a strong-lensing analysis of serendipitously discovered gravitational lensing arcs. 

In Section 2, we describe the observations, data reduction, and photometry. The detection of the giant arcs and our strong-lensing analysis are presented in Section 3. The concentration and mass of the cluster are estimated in Section 4. In section 5, we discuss the concentration and rarity of the cluster and conclude in Section 6.

We report magnitudes in the AB system. We assume a flat $\Lambda$CDM cosmology with $H_0=70$ km s$^{-1}$ Mpc$^{-1}$ and $\Omega_m$ = 0.3.

\section{Observations and Data Reduction} \label{sec:obs}

Observations of XLSSC~122 were carried out with JWST on 10 August 2024 (Program ID: 3950; PI: K. Finner). The primary purpose of the imaging was for weak-lensing analysis and the choices made for filters and integration times were based on our experience with infrared systematics \citep{2023bfinner} and our past weak-lensing analysis of the galaxy cluster SMACS J0723.3--7327 with JWST \citep{2023afinner}. 

JWST NIRCam can simultaneously image with a short- and long-wavelength filter, with this observation pairing F090W~\&~F277W and F200W~\&~F356W. We performed a MEDIUM8 readout pattern with 9 groups and 1 integration for a total exposure time of 7559~s for the F090W~\&~F277W pair. The pairing of F200W~\&~F356W was obtained in the SHALLOW8 readout pattern with 10 groups and 1 integration for a total exposure time of 4209~s. The cluster BCG was placed slightly off center from the middle of Module A so that the diffuse light surrounding it would not be severely impacted by the detector gaps. We reduced the observations with the JWST data reduction pipeline \citep{2025jwst_pipeline} and corrected systematic effects in the images with the YOUNG (Yonsei Observable UNiverse Group) JWST calibration pipeline\footnote{https://github.com/zpscofield/young-jwstpipe} \citep{2025scofield}. The individual observations were then stacked into mosaic images for each filter using the square kernel with a pixfrac of 0.75 and a final pixel scale of $0.02''$. The color image generated from the mosaics is presented in Fig. \ref{fig:xlssc122_image}. The \href{https://kylefinner.github.io/xlssc122}{JWST mosaics}\footnote{https://kylefinner.github.io/xlssc122} are publicly available. 

Photometric measurements were made with SExtractor \citep{1996sextractor} in dual-image mode. The detection image was created by weight-averaging the four mosaic images into a deep co-added image. The SExtractor parameters \texttt{DEBLEND\_MINAREA} and \texttt{DETECT\_THRESH} were set to 5 and 1.5, respectively. Although the high resolution of JWST is among the best for deblending galaxies, some galaxies require further deblending. We found that setting the deblending parameters \texttt{DEBLEND\_NTHRESH} and \texttt{DEBLEND\_MINCONT} to 32 and 0.005 was best for our purposes of separating the lensed images from the BCG and intracluster light. The $5\sigma$ point-source limiting magnitudes of the imaging are 29.2, 29.6, 30.9, and 30.7 for F090W, F200W, F277W, and F356W, respectively. The JWST photometry was combined with existing HST imaging in F814W, F105W, and F140W to derive photometric redshifts from 7-band photometry with {\tt eazy-py} \citep{2008brammer}.

Strong-lensing analysis can benefit from recent advancements in machine learning techniques. To search for additional multiple lensed images, we employed a Transformer architecture that deconvolves and denoises images. This architecture generated a denoised image with a signal-to-noise level boosted by 4. The denoised color image is shown in the lower left panel of Fig. \ref{fig:xlssc122_image}. The technique applied to denoise the image is fully described in \cite{2024park} and \cite{2025scofield}.

\begin{figure}[ht]
    \centering
    \includegraphics[width=\linewidth]{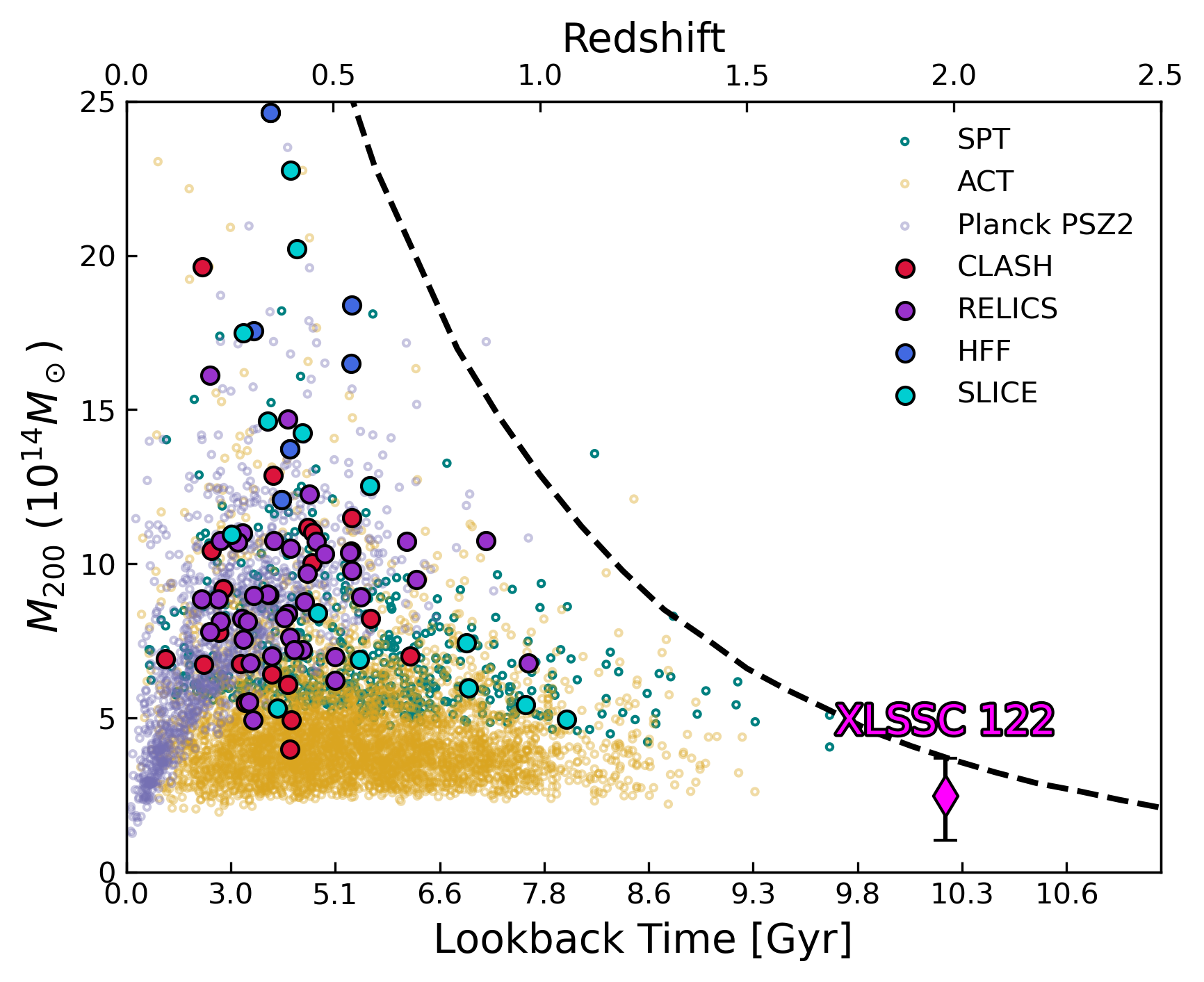}
    \caption{Galaxy cluster mass as a function of redshift for the SPT, ACT, and Planck surveys (open circles). Filled circles mark the galaxy clusters that have strong-lensing analyses from the CLASH, HFF, RELICS, and SLICE surveys. The black dashed line is the 95\% full-sky exclusion curve that separates expected (below) versus unexpected (above) cluster masses in the Universe. XLSSC~122 (mass from ACT) straddles the exclusions curve and sits at a substantially higher redshift than the galaxy clusters in the previous strong-lensing surveys.}
    \label{fig:mass_function}
\end{figure}
\section{Strong-Lensing Images and Analysis}\label{sec:SL_analysis}

\begin{figure*}[ht!]
    \centering
    \includegraphics[width=\linewidth]{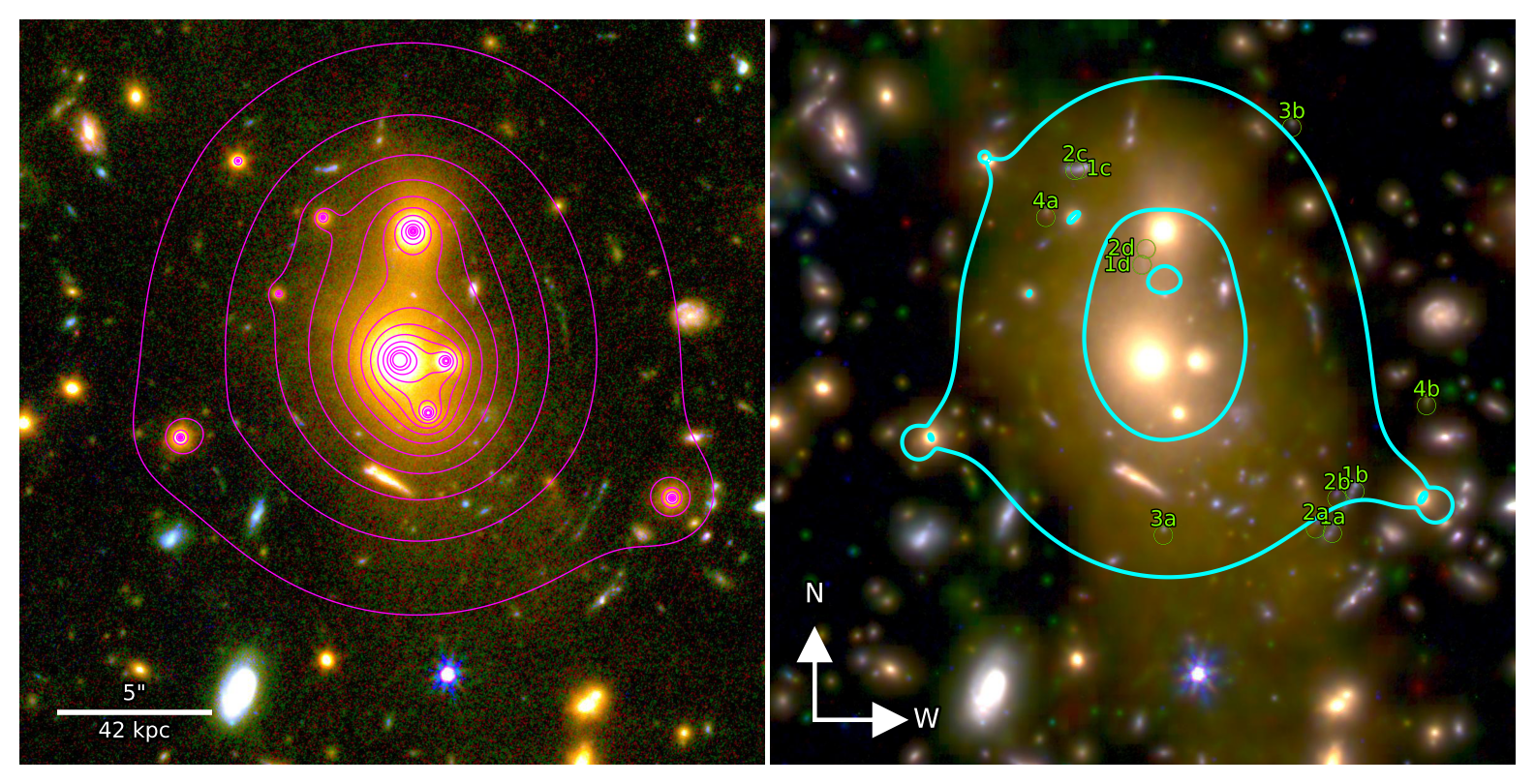}
    \caption{Left: Our parametric strong-lensing convergence map (magenta contours) overplotted on the color image of XLSSC~122. Right: Strong-lensing critical curves (cyan contours) for a source at $z=3.8$ plotted over the denoised color image. Multiple-image groups are marked with open circles and labeled with their identifier. }
    \label{fig:mass_recon}
\end{figure*}
In our deep JWST imaging of XLSSC~122, we made the remarkable discovery of giant arcs formed by the strong gravitational lensing effect at $z\mytilde2$ (presented in Fig. \ref{fig:xlssc122_image}). Two giant arcs (arcs 1 and 2) are found to the southwest of the brightest cluster galaxy (BCG). The giant arcs are similar in size, with a length-to-width ratio \mytilde20, and lie in close proximity on the sky. Two additional arc candidates are found in the imaging with one (arc 3) situated west of the BCG, with a length-to-width ratio similar to arcs 1 and 2, and a less prominent candidate (arc 4) to the east. These are the highest-redshift cluster-lensed giant arcs known to date, exceeding the previous record from the $z = 1.75$ cluster IDCS J1426.5+3508 \citep{2012gonzalez}. 

Fig. \ref{fig:mass_function} further puts the detection of the strong-lensing effect from XLSSC~122 into context. The figure shows the mass and redshift relation of galaxy clusters that have been detected in the Sunyaev Zel'dovich surveys (small circles) and highlights the large programs of strong-lensing analyses of galaxy clusters (big circles). For the cluster surveys, the most distant strong-lensing studies have been around a redshift of unity. The detection of strong-lensing arcs in XLSSC~122 at $z=1.98$ provides a new high--$z$ frontier for galaxy cluster studies. 

We use arcs 1 and 2, combined with additional lensed images (see Table \ref{table:images}), to perform a strong-lensing analysis of XLSSC~122. For brevity, we focus on the pertinent details of our analysis here and refer the reader to excellent review papers for theoretical backgrounds \citep[e.g.,][]{2010treu, 2024natarajan}. 

A strong gravitational lensing analysis uses the positions of the multiple images and distances to the lens and sources to develop a 2D model called the convergence ($\kappa$), which is a dimensionless map that can be calibrated to the projected mass density. In this work, we have modeled the mass distribution of the highest redshift cluster that exhibits strong lensing to date. To achieve that goal, we used the Lenstronomy \citep{2018birrer, 2021birrer} tools to implement a parametric modeling technique that uses prior knowledge of the cluster galaxies to set the positions of dark matter halos. This technique reduces the number of free parameters and improves constraints on the mass distribution when lensed images are sparse. 

We set the initial conditions for the model by identifying nine cluster galaxies within a projected distance of 100~kpc from the BCG, which are likely the most dominant contributors to the strong-lensing signal. At the location of each galaxy, we placed a Navarro--Frenk--White \citep[NFW;][]{1997navarro} halo with a fixed centroid. Lenstronomy parametrizes the NFW halos by their scale radius and deflection angle at the scale radius, which can be transformed to the mass and concentration of the halos. To account for the global mass distribution, we initialized a second NFW halo at the BCG position and allowed its centroid to move freely within a flat prior 2\arcsec~by~2\arcsec. The photometric redshifts we derived (in Section 2) for the multiple images are uncertain because of the limited photometric coverage. Rather than rely on the photometric redshifts for our strong-lensing analysis, we determined the average redshift for each multiple image group (Table \ref{table:images} $z$--phot. column), used them as the initial conditions, but allowed the redshifts to be fit (Table \ref{table:images} $z$--conv. column).

The groups of lensed images were provided to the algorithm and ray-tracing was used to derive the source-plane positions. We optimized the model by $\chi^2$ minimizing the projected distance between source-plane positions of the image groups. For the best-fit model, the source-plane rms scatter was $0.3''$.  The optimal convergence map and the critical curves of the model are shown in Fig. \ref{fig:mass_recon}.

To assess the uncertainty of the convergence map introduced by the poorly constrained redshifts of the lensed images, we performed a Monte Carlo sampling of the redshift range $2.1 < z < 10.0$.  For each multiple image group, a redshift was drawn from the range and fixed, while the NFW halos were optimized in the fitting procedure. The resampling and fitting procedure was repeated 1000 times and an rms convergence map was derived.

\section{Results} \label{sec:results}

\subsection{Radial Projected Mass Density}\label{mass_est}
Strong lensing tightly constrains the projected density of cluster cores. We converted the convergence map from our strong-lensing analysis to a projected mass density map by the relation $\Sigma = \kappa\Sigma_c(z)$, where $\Sigma_c(z)$ is the critical surface density at the redshift of the cluster. The critical surface density is defined as $\Sigma_c(z)=c^2D_s/(4\pi G D_l D_{ls}$), where $c$ is the speed of light, $G$ is the gravitational constant, and $D$ is the angular diameter distance to lens ($l$), source ($s$), and from lens to source ($ls$). 

\begin{figure}
    \centering
    \includegraphics[width=\linewidth]{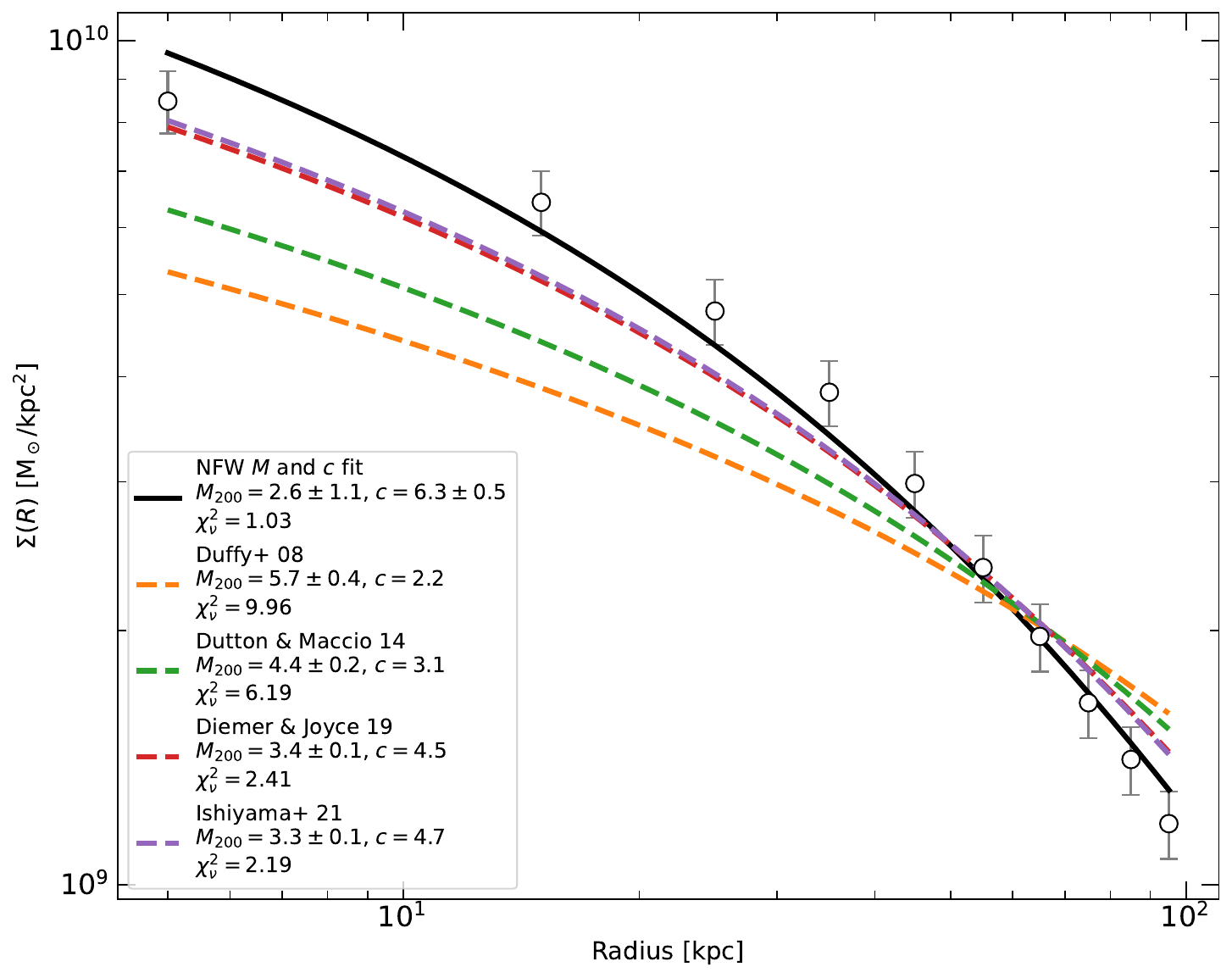}
    \caption{Radial mass density profile (black circles) from the strong-lensing model with best-fit NFW relation (black curve). Mass-concentration relations (dashed curves) from literature are fit to the radial profile. The mass-concentration relations are unable to reproduce the radial profile of XLSSC~122, cause an over-estimation of the mass, and an under-estimation of the concentration. Masses are reported in units of $10^{14}$ M$_\odot$. Reduced chi-square $\chi_\nu^2$ are presented for each model.}
    \label{fig:radial_profile}
\end{figure}

We determined the radial profile (black circles in Fig. \ref{fig:radial_profile}) by azimuthally averaging the projected mass density map with the center chosen to be the BCG. The uncertainty for each bin was extracted from the rms map that was derived from the 1000 convergence map realizations. The radial measurements were made to a maximum radius of 100 kpc, which is approximately the projected distance of the farthest lensed image. The radial measurements reveal an extremely cuspy inner core. We fit an NFW profile to the binned data following the formalism of \cite{2000wright}. The NFW profile fit to the inner 100 kpc (black curve in Figu. \ref{fig:radial_profile})  returns a concentration of $6.3\pm0.5$. This is an exceptionally high concentration for a dark matter halo at $z=1.98$. The exceptional concentration is further discussed in Subsection \ref{sec:conc}. 

Since strong lensing most strongly constrains the center of the cluster, we integrated the projected mass density map from the BCG location to a radius of 100~kpc. The cumulative mass within a radius of 100 kpc is $M$($R<$100 kpc) = $6.5\pm0.7\times10^{13}$ M$_\odot$, where the uncertainty is dominated by the scatter of the 1000 convergence map realizations that account for the uncertainty in source redshifts. 

It is also important to constrain the total cluster mass, commonly defined within a radius where the average density is 200 times the critical density of the universe. This requires extrapolation of the model to radii that are larger than the strong-lensing constraint but it has been shown to be a reliable method for predicting the total mass \citep{2025maraboli}. An NFW fit to the radial profile results in a mass of $M_{200c} = 2.6\pm1.1\times10^{14}$ M$_\odot$. This mass estimate is in agreement with the mass estimates that have come from the X-ray, weak lensing, and SZ measurements. For example, \cite{2024vanmar} estimate the mass as $M_{500c} = 1.66_{-0.20}^{+0.23}\times10^{14}$ M$_\odot$.  Furthermore, it highlights that XLSSC~122 is indeed a massive cluster that is seen at the epoch of the Universe where galaxy clusters are thought to first form. Further details into the nature of XLSSC~122 will be presented in an upcoming weak-lensing analysis (Scofield et al. in prep.).

\section{Discussion} \label{sec:discussion}

\subsection{High Concentration}\label{sec:conc}
Our strong-lensing model suggests that the concentration of XLSSC~122 is exceptionally high ($6.3\pm0.5$ at $z=1.98$). First, we put the concentration into the context of expected mass-concentration relations for dark matter halos of galaxy cluster scale ($10^{14}$ M$_\odot$). Then, we discuss how a high concentration at $z=1.98$ could be achieved.

Fig. \ref{fig:radial_profile} demonstrates that XLSSC~122 has an exceptionally high concentration compared to predictions from cosmological simulations. We fit a variety of mass-concentration relations from literature \citep{2008duffy, 2014dutton, 2019diemer, 2021ishiyama} and report their masses, concentrations, and reduced chi-squares in the figure. It is apparent that none of the mass-concentration relations fit the projected mass density well. The models predict a lower concentration than the concentration fit that we performed in Section \ref{sec:results}. The model concentrations range from $c=2.2$ (Duffy) to $c=4.7$ (Ishiyama), which are $8.2\sigma$ and $3.2\sigma$ lower than our best-fit concentration, respectively. These lower concentrations also suggest a higher total mass.

A high concentration can be the effect of a triaxial halo observed along the line of sight. However, there is evidence that supports the halo not being triaxial along the line of sight. First, \cite{2024vanmar} showed that the pressure peak, measured from the SZ, is offset to the south of the BCG and X-ray peak. For a line-of-sight triaxial cluster or a line-of-sight merger, it is not expected that the pressure peak would be offset from the X-ray brightness peak. Second, the mass distribution stretches significantly in the plane of sky, suggesting that it is unlikely to be stretched along the line of sight as well. The weak--lensing mass distribution \citep[shown in Fig. 2 of][]{2025kim} shows tentative ($<3\sigma$) evidence that the mass distribution is elongated in the plane of the sky. Using the JWST observations, Z. Scofield et al. (in prep.) shows a significant detection of an elongated mass distribution and combines it with multiwavelength observations to describe the dynamical state of the cluster. In addition, the JWST observations detect an intracluster light (see Fig. \ref{fig:xlssc122_image} or \ref{fig:mass_recon}) that is extremely extended along the same axis as the weak-lensing mass distribution. The intracluster light provides an additional indicator to the past merger history of the cluster and will be investigated in H. Joo et al. (in prep.). However, XLSSC~122 is observed at an epoch where the first galaxy clusters are forming and is a regime that has very few galaxy clusters studied. Thus, we cannot conclusively state that XLSSC~122 is not significantly elongated along the line of sight. A detailed spectroscopic analysis of the cluster galaxies is a better approach to discern the line-of-sight distribution of matter.

The high concentration has significant ramifications for the assembly history of XLSSC~122 and is an indication that the halo collapsed early \citep{2001bullock, 2002wechsler}. \cite{2021noordeh} has already shown that a population of `bulge-like' quiescent galaxies in the cluster are on average 65\% larger than their field counterparts at $z\sim2$, which suggests that the cluster member galaxies have experienced an accelerated size evolution relative to the field at $z>2$. The same accelerated size evolution for the galaxies may have impacted the global dark matter halo, accelerating its growth, and leading to the high concentration. 

Alternatives and/or variations of the $\Lambda$CDM model have gained momentum with the JWST discoveries of high--$z$ galaxies. One avenue to the early formation of massive halos is through early dark energy \citep[see][for examples on how early dark energy impacts galaxy formation]{2024shen}. Early dark energy facilitates the earlier formation of dark matter halos, which provides the scaffolding for rapid galaxy formation. The more massive halos accrete the surrounding gas and dark matter faster, leading to more mergers and more rapid growth. Early dark energy has also been discussed in the context of solving the arcs statistics problem \citep{2013meneghetti}, of which XLSSC~122 is a compelling example due to its prominent giant arcs. 

\subsection{Rarity} \label{sec:rarity}

In Fig. \ref{fig:mass_function}, we plot the masses of galaxy clusters from the \cite{2016planck}, South Pole Telescope \citep[SPT;][]{2020bleem}, and Atacama Cosmology Telescope \citep[ACT;][]{2021hilton} surveys. Clusters from strong-lensing studies of Hubble Frontier Fields \citep[HFF;][]{2019coe}, Reionization Clusters Survey \citep[RELICS;][]{2017lotz}, and Strong Lensing and Cluster Evolution \citep[SLICE;][]{2025cerny} are marked with solid circles. We draw an exclusion curve that denotes the 95\% upper bound of galaxy cluster mass at a given redshift. This exclusion curve is sensitive to the $\sigma_8$ parameter and the halo mass function (HMF) for which we use the Planck value $\sigma_8=0.816\pm0.009$ and the HMF of \cite{2016bocquet}. We integrate the HMF over the redshift and mass intervals of $1.98 < z < 10$ and $14.4 <$ log($M$) $< 16$ to determine the probability of detecting a cluster of XLSSC~122 mass or greater. The integration suggests that in the survey field of the XXL, only 0.12 clusters like XLSSC~122 should exist. This puts XLSSC~122 into the rare category of massive clusters at high redshift that are improbable detections such as SPT-CL J2040-4451 \citep{2017jee} and SpARCS J1049+56 \citep{2020finner}. Expanding to the full sky, 193 clusters of at least XLSSC~122 mass are expected.

\section{Conclusions} \label{sec:conclusion}
In deep JWST NIRCam imaging of the galaxy cluster XLSSC~122 ($z=1.98$), we have made the remarkable discovery of giant strong gravitational lensing arcs. We used the arcs and additional multiple images to perform a strong-lensing analysis of XLSSC~122 and constrain the projected mass density in the cluster core.

We derived a convergence map for the cluster using a parametric strong-lensing method and 4 systems of multiple images. The radial projected mass density profile was measured and found to be highly concentrated. Fitting an NFW model to the inner 100 kpc of the mass density profile, we measured a concentration of $6.3\pm0.5$. This concentration is found to be exceptionally high for a galaxy cluster at $z=1.98$.  We demonstrated that the exceedingly high concentration is not predicted by mass-concentration relations and we discussed potential avenues for the formation of a highly concentrated cluster at $z=1.98$.

We integrated the projected mass density map to find the mass within $100$ kpc of the BCG to be $6.5\pm0.7\times 10^{13}$ M$_\odot$. We fit an NFW model to the radial projected mass density profile and determine the total mass $M_{200c} = 2.6\pm1.1\times 10^{14}$ M$_\odot$. Our mass estimate is in agreement with previous studies of the cluster and reinforces that XLSSC~122 is a massive and rare cluster at an early stage of the universe. 

\section{Software and third party data repository citations} \label{sec:cite}
\emph{SExtractor} \citep{1996sextractor}; \emph{Colossus} \citep{2018diemer}; \emph{JWST Data Reduction} \citep{2025jwst_pipeline}; \emph{YOUNG JWST Pipeline} \citep{2025scofield}; \emph{Lenstronomy} \citep{2018birrer, 2021birrer}, \emph{eazy-py} \citep{2008brammer}.
%% Please use the acknowledgment and contribution environments. This will 
%% be anonomyized when the "anonymous" style option is used. 
\begin{acknowledgments}

S.C. acknowledges that this research was supported by the Basic Science Research Program through the National Research Foundation of Korea (NRF) funded by the Ministry of Education (No. RS-2024-00413036), and International Joint Research Grant by Yonsei Graduate School. B. Lee is supported by the National Research Foundation of Korea(NRF) grant funded by the Korea government(MSIT), No. NRF-2022R1C1C1008695. 

This work is based [in part] on observations made with the NASA/ESA/CSA James Webb Space Telescope. The data were obtained from the Mikulski Archive for Space Telescopes at the Space Telescope Science Institute, which is operated by the Association of Universities for Research in Astronomy, Inc., under NASA contract NAS 5-03127 for JWST. These observations are associated with program \#3950.

Support for program \#3950 was provided by NASA through a grant from the Space Telescope Science Institute, which is operated by the Association of Universities for Research in Astronomy, Inc., under NASA contract NAS 5-03127.

\end{acknowledgments}

\begin{contribution}
K. Finner,  S. Cha, and Z. Scofield performed the strong-lensing analysis. K. Finner wrote the majority of the manuscript. M. J. Jee provided expertise in gravitational lensing and contributed to the writing of the manuscript. Y. Lin performed grism data reduction and redshift estimation. T. Morishita and H. Joo estimated photometric redshifts. H. Park created the denoised image. A. Faisst, B. Lee, W. Wang, and R. Chary contributed to the JWST proposal and the writing of the manuscript. 
%%This section gives authors the space to recognize author contributions. The text inside this environment is NOT counted towards the total word quanta. At a minimum, manuscripts are expected to include this text:

%% But authors are expected to provide more specific details, e.g. 
%%
%%SC was responsible for writing and submitting the manuscript.
%%WWM came up with the initial research concept and edited the manuscript.
%%OTS obtained the funding and edited the manuscript.
%%EBF provided the formal analysis and validation. He also edited the manuscript.
%%GEH Supervised the undergraduates, wrote the software and administers the project github and Zenodo repositories.
%%
%% Authors can use the Contributor Role Taxonomy (CRediT) at
%% for ideas on how write a good statement tailored to their needs.

\end{contribution}

\facilities{HST(WFC3), JWST(NIRCam)}



\begin{table}[]
\centering
\caption{Lensed image groups.}
\label{table:images}
\begin{tabular}{lllll}
Image & R.A. & Dec.  & $z$--phot. & $z$--conv. \\%& \\ % $dz$ \\
\hline\hline
1a  & 34.4326008 & -3.7602935    & 3.6 & 3.5  \\% & 1.3    \\
1b  & 34.4323985   & -3.7599181    & 3.6 & 3.5 \\% & 1.3    \\
1c  & 34.4348622   & -3.7570362    & 3.6 & 3.5 \\% & 1.3    \\
1d  & 34.4343030   & -3.7578504    & 3.6 & 3.5 \\% & 1.3    \\
2a  & 34.4327508 & -3.7602543    & 4.3 & 2.9 \\% & 1.3  \\
2b  &  34.4325590  & -3.7599825     & 4.3 & 2.9\\% & 1.3  \\
2c  &  34.4349075  & -3.7570480     & 4.3 & 2.9\\% & 1.3  \\
2d  &  34.4342780  & -3.7577550     & 4.3 & 2.9\\% & 1.3  \\
3a  & 34.4341029   & -3.7602667    & 3.7 & 3.0 \\% & 2.0   \\
3b  & 34.4329571    & -3.7566203    & 3.7 & 3.0\\% & 2.0   \\
4a  & 34.4317481   & -3.7591029    & 2.9 & 4.5 \\% & 2.5  \\
4b  & 34.4351551   & -3.7574216    & 2.9 & 4.5 \\% & 2.5  \\

\end{tabular}
\end{table}

\bibliography{sample701}{}
\bibliographystyle{aasjournalv7}



\end{document}